\input amssym.def
\input epsf


\magnification=\magstephalf
\hsize=14.0 true cm
\vsize=19 true cm
\hoffset=1.0 true cm
\voffset=2.0 true cm

\abovedisplayskip=12pt plus 3pt minus 3pt
\belowdisplayskip=12pt plus 3pt minus 3pt
\parindent=1.0em


\font\sixrm=cmr6
\font\eightrm=cmr8
\font\ninerm=cmr9

\font\sixi=cmmi6
\font\eighti=cmmi8
\font\ninei=cmmi9

\font\sixsy=cmsy6
\font\eightsy=cmsy8
\font\ninesy=cmsy9

\font\sixbf=cmbx6
\font\eightbf=cmbx8
\font\ninebf=cmbx9

\font\eightit=cmti8
\font\nineit=cmti9

\font\eightsl=cmsl8
\font\ninesl=cmsl9

\font\sixss=cmss8 at 8 true pt
\font\sevenss=cmss9 at 9 true pt
\font\eightss=cmss8
\font\niness=cmss9
\font\tenss=cmss10

 at 12 true pt
\font\bigrm=cmr10 at 12 true pt
 at 12 true pt

 at 14 true pt
\font\Bigrm=cmr12 at 16 true pt
 at 14 true pt

\catcode`@=11
\newfam\ssfam

\def\tenpoint{\def\rm{\fam0\tenrm}%
    \textfont0=\tenrm \scriptfont0=\sevenrm \scriptscriptfont0=\fiverm
    \textfont1=\teni  \scriptfont1=\seveni  \scriptscriptfont1=\fivei
    \textfont2=\tensy \scriptfont2=\sevensy \scriptscriptfont2=\fivesy
    \textfont3=\tenex \scriptfont3=\tenex   \scriptscriptfont3=\tenex
    \textfont\itfam=\tenit                  \def\it{\fam\itfam\tenit}%
    \textfont\slfam=\tensl                  \def\sl{\fam\slfam\tensl}%
    \textfont\bffam=\tenbf \scriptfont\bffam=\sevenbf
    \scriptscriptfont\bffam=\fivebf
                                            \def\bf{\fam\bffam\tenbf}%
    \textfont\ssfam=\tenss \scriptfont\ssfam=\sevenss
    \scriptscriptfont\ssfam=\sevenss
                                            \def\ss{\fam\ssfam\tenss}%
    \normalbaselineskip=13pt
    \setbox\strutbox=\hbox{\vrule height8.5pt depth3.5pt width0pt}%
    \let\big=\tenbig
    \normalbaselines\rm}

\def\ninepoint{\def\rm{\fam0\ninerm}%
    \textfont0=\ninerm      \scriptfont0=\sixrm
                            \scriptscriptfont0=\fiverm
    \textfont1=\ninei       \scriptfont1=\sixi
                            \scriptscriptfont1=\fivei
    \textfont2=\ninesy      \scriptfont2=\sixsy
                            \scriptscriptfont2=\fivesy
    \textfont3=\tenex       \scriptfont3=\tenex
                            \scriptscriptfont3=\tenex
    \textfont\itfam=\nineit \def\it{\fam\itfam\nineit}%
    \textfont\slfam=\ninesl \def\sl{\fam\slfam\ninesl}%
    \textfont\bffam=\ninebf \scriptfont\bffam=\sixbf
                            \scriptscriptfont\bffam=\fivebf
                            \def\bf{\fam\bffam\ninebf}%
    \textfont\ssfam=\niness \scriptfont\ssfam=\sixss
                            \scriptscriptfont\ssfam=\sixss
                            \def\ss{\fam\ssfam\niness}%
    \normalbaselineskip=12pt
    \setbox\strutbox=\hbox{\vrule height8.0pt depth3.0pt width0pt}%
    \let\big=\ninebig
    \normalbaselines\rm}

\def\eightpoint{\def\rm{\fam0\eightrm}%
    \textfont0=\eightrm      \scriptfont0=\sixrm
                             \scriptscriptfont0=\fiverm
    \textfont1=\eighti       \scriptfont1=\sixi
                             \scriptscriptfont1=\fivei
    \textfont2=\eightsy      \scriptfont2=\sixsy
                             \scriptscriptfont2=\fivesy
    \textfont3=\tenex        \scriptfont3=\tenex
                             \scriptscriptfont3=\tenex
    \textfont\itfam=\eightit \def\it{\fam\itfam\eightit}%
    \textfont\slfam=\eightsl \def\sl{\fam\slfam\eightsl}%
    \textfont\bffam=\eightbf \scriptfont\bffam=\sixbf
                             \scriptscriptfont\bffam=\fivebf
                             \def\bf{\fam\bffam\eightbf}%
    \textfont\ssfam=\eightss \scriptfont\ssfam=\sixss
                             \scriptscriptfont\ssfam=\sixss
                             \def\ss{\fam\ssfam\eightss}%
    \normalbaselineskip=10pt
    \setbox\strutbox=\hbox{\vrule height7.0pt depth2.0pt width0pt}%
    \let\big=\eightbig
    \normalbaselines\rm}

\def\tenbig#1{{\hbox{$\left#1\vbox to8.5pt{}\right.\n@space$}}}
\def\ninebig#1{{\hbox{$\textfont0=\tenrm\textfont2=\tensy
                       \left#1\vbox to7.25pt{}\right.\n@space$}}}
\def\eightbig#1{{\hbox{$\textfont0=\ninerm\textfont2=\ninesy
                       \left#1\vbox to6.5pt{}\right.\n@space$}}}

\font\sectionfont=cmbx10
\font\subsectionfont=cmti10

\def\figurecaptionfont{\ninepoint}
\def\tablecaptionfont{\ninepoint}
\def\footnotefont{\eightpoint}


\newcount\equationno
\newcount\bibitemno
\newcount\figureno
\newcount\tableno

\equationno=0
\bibitemno=0
\figureno=0
\tableno=0


\footline={\ifnum\pageno=0{\hfil}\else
{\hss\rm\the\pageno\hss}\fi}


\def\section #1. #2 \par
{\vskip0pt plus .20\vsize\penalty-100 \vskip0pt plus-.20\vsize
\vskip 1.6 true cm plus 0.2 true cm minus 0.2 true cm
\global\def\equationlabel{#1}
\global\equationno=0
\leftline{\sectionfont #1. #2}\par
\immediate\write\terminal{Section #1. #2}
\vskip 0.7 true cm plus 0.1 true cm minus 0.1 true cm
\noindent}


\def\subsection #1 \par
{\vskip0pt plus 0.8 true cm\penalty-50 \vskip0pt plus-0.8 true cm
\vskip2.5ex plus 0.1ex minus 0.1ex
\leftline{\subsectionfont #1}\par
\immediate\write\terminal{Subsection #1}
\vskip1.0ex plus 0.1ex minus 0.1ex
\noindent}


\def\appendix #1. #2 \par
{\vskip0pt plus .20\vsize\penalty-100 \vskip0pt plus-.20\vsize
\vskip 1.6 true cm plus 0.2 true cm minus 0.2 true cm
\global\def\equationlabel{\hbox{\rm#1}}
\global\equationno=0
\leftline{\sectionfont Appendix #1. #2}\par
\immediate\write\terminal{Appendix #1. #2}
\vskip 0.7 true cm plus 0.1 true cm minus 0.1 true cm
\noindent}



\def\equation#1{$$\displaylines{\qquad #1}$$}
\def\enum{\global\advance\equationno by 1
\hfill\llap{(\equationlabel.\the\equationno)}}

\def\next#1{\cr\noalign{\vskip#1}\qquad}


\def\ifundefined#1{\expandafter\ifx\csname#1\endcsname\relax}

\def\ref#1{\ifundefined{#1}?\immediate\write\terminal{unknown reference
on page \the\pageno}\else\csname#1\endcsname\fi}

\newwrite\terminal
\newwrite\bibitemlist

\def\bibitem#1#2\par{\global\advance\bibitemno by 1
\immediate\write\bibitemlist{\string\def
\expandafter\string\csname#1\endcsname
{\the\bibitemno}}
\item{[\the\bibitemno]}#2\par}

\def\beginbibliography{
\vskip0pt plus .15\vsize\penalty-100 \vskip0pt plus-.15\vsize
\vskip 1.2 true cm plus 0.2 true cm minus 0.2 true cm
\leftline{\sectionfont References}\par
\immediate\write\terminal{References}
\immediate\openout\bibitemlist=biblist
\frenchspacing\parindent=1.8em
\vskip 0.5 true cm plus 0.1 true cm minus 0.1 true cm}

\def\endbibliography{
\immediate\closeout\bibitemlist
\nonfrenchspacing\parindent=1.0em}


\def\figurecaption#1{\global\advance\figureno by 1
\narrower\figurecaptionfont
Fig.~\the\figureno. #1}

\def\tablecaption#1{\global\advance\tableno by 1
\vbox to 0.5 true cm { }
\centerline{\tablecaptionfont%
Table~\the\tableno. #1}
\vskip-0.4 true cm}

\def\thicktablerule{\hrule height1pt}
\def\thintablerule{\hrule height0.4pt}

\tenpoint

\immediate\openin\bibitemlist=biblist
\ifeof\bibitemlist\immediate\closein\bibitemlist
\else\immediate\closein\bibitemlist
\input biblist \fi


\def\thismonth{\ifcase\month\or
January\or February\or March\or April\or May\or June\or
July\or August\or September\or October\or November\or December\fi}



\def\rmd{{\rm d}}
\def\rmD{{\rm D}}
\def\rme{{\rm e}}
\def\rmO{{\rm O}}


\def\gz{{\Bbb Z}}


\def\proof{\noindent{\sl Proof:}\kern0.6em}

\def\frac#1#2{\hbox{$#1\over#2$}}
\def\dual{\mathstrut^*\kern-0.1em}

\def\lvec#1{\setbox0=\hbox{$#1$}
    \setbox1=\hbox{$\scriptstyle\leftarrow$}
    #1\kern-\wd0\smash{
    \raise\ht0\hbox{$\raise1pt\hbox{$\scriptstyle\leftarrow$}$}}
    \kern-\wd1\kern\wd0}
\def\rvec#1{\setbox0=\hbox{$#1$}
    \setbox1=\hbox{$\scriptstyle\rightarrow$}
    #1\kern-\wd0\smash{
    \raise\ht0\hbox{$\raise1pt\hbox{$\scriptstyle\rightarrow$}$}}
    \kern-\wd1\kern\wd0}


\def\nabstar#1{{\nabla\kern0.5pt\smash{\raise 4.5pt\hbox{$\ast$}}
               \kern-5.5pt_{#1}}}

\def\drvstar#1{{\partial\kern0.5pt\smash{\raise 4.5pt\hbox{$\ast$}}
               \kern-6.0pt_{#1}}}

\def\ldrvstar#1{{\lvec{\,\partial}\kern-0.5pt\smash{\raise 4.5pt\hbox{$\ast$}}
               \kern-5.0pt_{#1}}}


\def\fm{{\rm fm}}




\def\diracstar#1#2{
    \setbox0=\hbox{$\gamma$}\setbox1=\hbox{$\gamma_{#1}$}
    \gamma_{#1}\kern-\wd1\kern\wd0
    \smash{\raise4.5pt\hbox{$\scriptstyle#2$}}}


\def\tr{{\rm tr}}
\def\Tr{{\rm Tr}}
\def\Ad{{\rm Ad}\kern0.1em}


\def\sc{{\gamma}}
\def\tm{{\Bbb T}}
\def\pvc{{\Bbb P}}
\def\pqq{{\Bbb P}_{\bf 3\otimes3^{\ast}}(x,y)}
\def\rbar{\bar{r}}
\def\rtilde{\tilde{r}}
\def\gbar{\bar{g}}
\rightline{CERN-TH/2002-138}
\rightline{MPI-PhT/2002-24}

\vskip 2.0 true cm 
\centerline{\Bigrm Quark confinement and the bosonic string}
\vskip 0.6 true cm
\centerline{\bigrm Martin L\"uscher\kern1pt%
\footnote{${\vrule height7.0pt depth1.5pt width0pt}^{\ast}$}
{\footnotefont%
\hskip-1ex On leave from Deutsches Elektronen-Synchrotron DESY, 
D-22603 Hamburg, Germany}
}
\vskip1ex
\centerline{\it CERN, Theory Division} 
\centerline{\it CH-1211 Geneva 23, Switzerland}
\vskip 0.4 true cm
\centerline{\bigrm Peter Weisz}
\vskip1ex
\centerline{\it Max-Planck-Institut f\"ur Physik}
\centerline{\it D-80805 Munich, Germany}
\vskip 0.8 true cm
\thintablerule
\vskip 2.0ex
\ninepoint
\leftline{\bf Abstract}
\vskip 1.0ex\noindent
Using a new type of simulation algorithm 
for the standard SU(3) lattice gauge theory that yields results
with unprecedented precision,
we confirm the presence of a $\sc/r$ correction
to the static quark potential at large distances $r$,
with a coefficient $\sc$ as predicted by the bosonic string theory.
In both three and four dimensions,
the transition from perturbative to 
string behaviour is evident from the data and
takes place at surprisingly small distances.
\vskip 2.0ex
\thintablerule

\tenpoint

\section 1. Introduction

The idea that the pure SU(3) gauge theory 
(and all other non-abelian gauge theories) might be closely  
related to some kind of string theory has been around for a long time. 
A comparatively direct connection derives from the
observation that the gauge field generated by two 
widely separated static quarks
appears to be squeezed into a flux tube.
Such flux tubes may be expected to behave like 
strings with fixed ends, at least in the limit
where they are very much longer than wide.

On a more formal level it has been suggested
that the expectation values of large Wilson loops are matched by
the corresponding amplitudes of an
effective bosonic string theory [\ref{Nambu}].
Assuming this to be the case,
the static quark potential $V(r)$ can be
shown to have an asymptotic expansion
\equation{
  V(r)=
  \sigma r+\mu+\sc/r+\rmO(1/r^2),
  \qquad \sc=-{\pi\over 24}\left(d-2\right),
  \enum
}
at large distances $r$,
where $\sigma$ denotes the string tension,
$\mu$ a regularization-dependent mass
and $d$ the dimension of space-time
[\ref{WKB},\ref{UniversalTerm}].
The $\sc/r$ correction in this formula is a quantum effect
that is characteristic of the relativistic bosonic string. 
It is universal in the sense that the value of $\sc$
is the same for a large class of string actions.

In principle the validity of eq.~(1.1) 
can be checked by putting the theory
on a lattice and by calculating $V(r)$ through numerical simulation.
The statistical and systematic errors in these computations
are, however, rapidly
increasing at large distances, and
it is then practically impossible to clearly separate the $\sc/r$ 
correction from the other terms 
(if the standard simulation algorithms are used).
As a result, we would find it difficult
to claim, on the basis of the published simulation data alone, 
that the expansion holds with the quoted value of $\sc$.

There is nevertheless some support from lattice gauge theory
for the string model.
It has been noted, for example, that 
the data for the potential at distances between $0.4$ and $0.8$ fm
agree with eq.~(1.1) within small errors [\ref{SilviaRainer}].
Another numerical study that we wish to mention here
is the work of Caselle et al.~[\ref{CaselleEtAl}]
on the confinement phase of the $\gz_2$ gauge theory in three dimensions.
Highly efficient simulation techniques
are available in this case, and using these it was possible
to confirm that the expectation values of large rectangular Wilson loops
are matched by the string theory amplitudes 
to a precision where the subleading string effects
can be clearly identified.

Last year we proposed a new type of simulation algorithm for 
the SU(3) gauge theory that leads to an exponential reduction 
of the statistical errors in calculations of 
the Polyakov loop correlation function [\ref{MultiLevel}].
Similar to the expectation values of Wilson loops,
the latter can serve as a probe for string effects
[\ref{AmbjornEtAl},\ref{deForcrandEtAl}], and
the onset of string behaviour 
may in fact be easier to observe in this way [\ref{LuciniTeper}].
In the present paper, though,
the correlation function only appears
at an intermediate stage. 
The computations are well adapted to the new algorithm,
and the results that we obtain for the force 
$V'(r)$ and the second derivative $V''(r)$
are sufficiently precise for the
question raised above to be addressed.

\section 2. Polyakov loop correlation function

In this section we recall some
well-known facts about the Polyakov loop correlation function
in the SU(3) gauge theory.
The notation is the same as in ref.~[\ref{MultiLevel}],
except that we now consider the theory also in three dimensions.

\subsection 2.1 Definition

We set up the lattice theory in the standard manner
on a 4-dimensional hypercubic or 3-dimensional cubic lattice with
spacing $a$, time-like extent $T$ and spatial size $L$.
Periodic boundary conditions are imposed in all directions, and
for the gauge field action we take the Wilson plaquette action
$S[U]$ with bare gauge coupling $g_0$~[\ref{Wilson}].

For any given gauge field configuration $U(x,\mu)$,
the Polyakov loop 
\equation{
  P(x)=\tr\left\{
  U(x,\mu)U(x+a\hat{\mu},\mu)\ldots U(x+(T-a)\hat{\mu},\mu)\right\}_{\mu=0}
  \enum
}
is the trace of the Wilson line that passes through $x$ and 
that winds once around the world along the time axis
($\hat{\mu}$ denotes the unit vector in direction $\mu$).
We are then interested in the correlation function
\equation{
  \langle P(x)^{\ast} P(y)\rangle=
  {1\over{\cal Z}}\int\rmD[U]\,P(x)^{\ast} P(y)\,\rme^{-S[U]},
  \qquad
  \rmD[U]=\prod_{x,\mu}\rmd U(x,\mu),
  \enum
}
where
${\cal Z}$ is a normalization factor such that
$\langle 1\rangle=1$ and $\rmd U(x,\mu)$ stands for 
the normalized invariant measure on SU(3).

\subsection 2.2 Spectral representation

An important property of 
the Polyakov loop correlation function is that it can be written as a
ratio of partition functions.
The transfer matrix formalism must
be invoked to show this,
but the derivation is then fairly straightforward
(for an introduction to the subject, see 
refs.~[\ref{TransferI}--\ref{TransferIV}] for example).

First recall that the quantum mechanical states of the theory
are represented by wave functions 
on the space of all spatial lattice gauge fields $U$ at time $x_0=0$. 
The transfer matrix in the temporal gauge, $\tm\equiv\rme^{-{\Bbb H}a}$,
is a bounded hermitian operator that acts on these functions.
In this framework the gauge-invariant wave functions describe
the physical states in the vacuum sector,
and the wave functions with two colour indices 
that transform according to
\equation{
  \psi_{\alpha\beta}[U^{\Lambda}]=\Lambda(x)_{\alpha\gamma}
  [\Lambda(y)_{\beta\delta}]^{\ast}
  \psi_{\gamma\delta}[U]
  \enum
}
under gauge transformations $\Lambda$
(where $U^{\Lambda}$ denotes the gauge transform of $U$)
describe states
with a static quark at the point $x$ and a static antiquark at $y$.

If we introduce the projector $\pvc$ to the subspace 
of gauge-invariant wave functions, and similarly the 
projector $\pqq$ to the subspace of wave functions (2.3),
the partition functions in these sectors are given by
\equation{
  {\cal Z}=
  \Tr\left\{
  {\pvc}
  \,\rme^{-{\Bbb H}T}\right\},
  \enum
  \next{2.3ex}
  {\cal Z}_{\bf 3\otimes3^{\ast}}(x,y)=
  \frac{1}{9}\kern1pt\Tr\left\{
  \pqq
  \,\rme^{-{\Bbb H}T}\right\}.
  \enum
}
A conventional normalization factor has been included here to 
compensate for the trivial degeneracy of the states in the
multiplets (2.3).
The partition function (2.5) is thus effectively a sum over multiplets.

By inserting the explicit expression for 
the transfer matrix, both partition functions can be rewritten in 
the form of functional integrals on the $d$-dimensional lattice.
Using an SU(3) character relation, it is then possible to 
establish the identity
\equation{
  \left\langle P(x)^{\ast}P(y)\right\rangle=
  {{\cal Z}_{\bf 3\otimes3^{\ast}}(x,y)\over{\cal Z}}.
  \enum
}
The spectrum of the transfer matrix is
purely discrete in finite volume,
and since the ground state in the vacuum sector is non-degenerate
[\ref{TransferII},\ref{TransferIII}],
an immediate consequence of this representation is that
\equation{
  \left\langle P(x)^{\ast}P(y)\right\rangle=
  \sum_{n=0}^{\infty}w_n\kern1pt\rme^{-E_nT}
  \enum
}
for some positive energies $E_n$ and {\it integral}\/ weights $w_n$.

\subsection 2.3 Relation to the potential $V(r)$

We shall only consider the potential along the 
lattice axes and thus set $y=x+r\hat{\mu}$, $\mu=1$, in the following.
If we arrange the energy values in eq.~(2.7) in ascending order,
it is clear that $E_0$ is the lowest energy
(above the ground state energy)
in the sector with the static quarks.
In other words, $E_0$ coincides with the static quark potential $V(r)$.

The associated eigenstate can be shown to be non-degenerate
at strong coupling, and 
in our numerical studies of the Polyakov loop correlation function
we have always found that $w_0=1$. It is usually fairly easy to obtain this
result, since $w_0$ is known to be a positive integer.
We thus conclude that
\equation{
   V(r)=-{1\over T}\ln\left\langle P(x)^{\ast}P(y)\right\rangle
   +\varepsilon,
   \enum
}
where 
\equation{
   \varepsilon={1\over T}\left\{w_1\rme^{-\Delta E T}+\ldots\right\},
   \qquad
   \Delta E=E_1-E_0.
   \enum
}
In particular, no fits are required 
to extract the potential 
from the Polyakov loop correlation function once the latter
has been computed at large values of $T$, where the error
$\varepsilon$ is negligible.

\section 3. Bosonic string theory

The following paragraphs contain a brief description
of the string theory representation of the Polyakov loop
correlation function. 
It is included here mainly
for the reader's convenience, but also to prepare the 
ground for the analysis of the simulation data.

\subsection 3.1 Effective action

In the bosonic string theory 
the Polyakov loop correlation function is 
formally given by a functional integral 
over all two-dimensional surfaces (world-sheets) that
are bounded by the loops [\ref{Nambu}].
When both $T$ and $r$ are large with respect to the scale 
set by the string tension $\sigma$,
the integral is dominated by the surface with the minimal area
and can be calculated by expanding 
the integrand around this surface [\ref{WKB}].

The important degree of freedom in this situation is the deviation 
of the fluctuating surface from the minimal surface 
in the directions orthogonal to the latter (fig.~1).
If we parameterize the minimal surface through
two cartesian coordinates $z_a$ ($a=0,1$) in the obvious way,
the deviation is a vector field
\equation{
  h(z)=\left(0,0,h_2(z),\ldots,h_{d-1}(z)\right)
  \enum
}
with $d-2$ non-zero transverse components that satisfy Dirichlet boundary
conditions at $z_1=0$ and $z_1=r$.
The associated effective action may depend on how precisely
the underlying bosonic string theory was set up,
but for symmetry reasons it has to be of the form 
\equation{
   S_{\rm eff}=
   \int_0^T\int_0^r\rmd z_0\rmd z_1
   \left\{\frac{1}{2}\partial_ah\partial_ah+\ldots\right\},
   \enum
}
where the ellipses stand
for higher-dimensional interaction terms that are Lorentz-invariant 
polynomials in the derivatives of $h$ [\ref{UniversalTerm}].
In this context the field $h$ is counted as dimensionless
and it is, therefore, natural to assume from the beginning that
the displacement of the fluctuating surface in physical units 
is $h/\sqrt{\sigma}$ rather than $h$.

\topinsert
\vbox{
\vskip0.0cm
\epsfxsize=6.0cm\hskip3.0cm\epsfbox{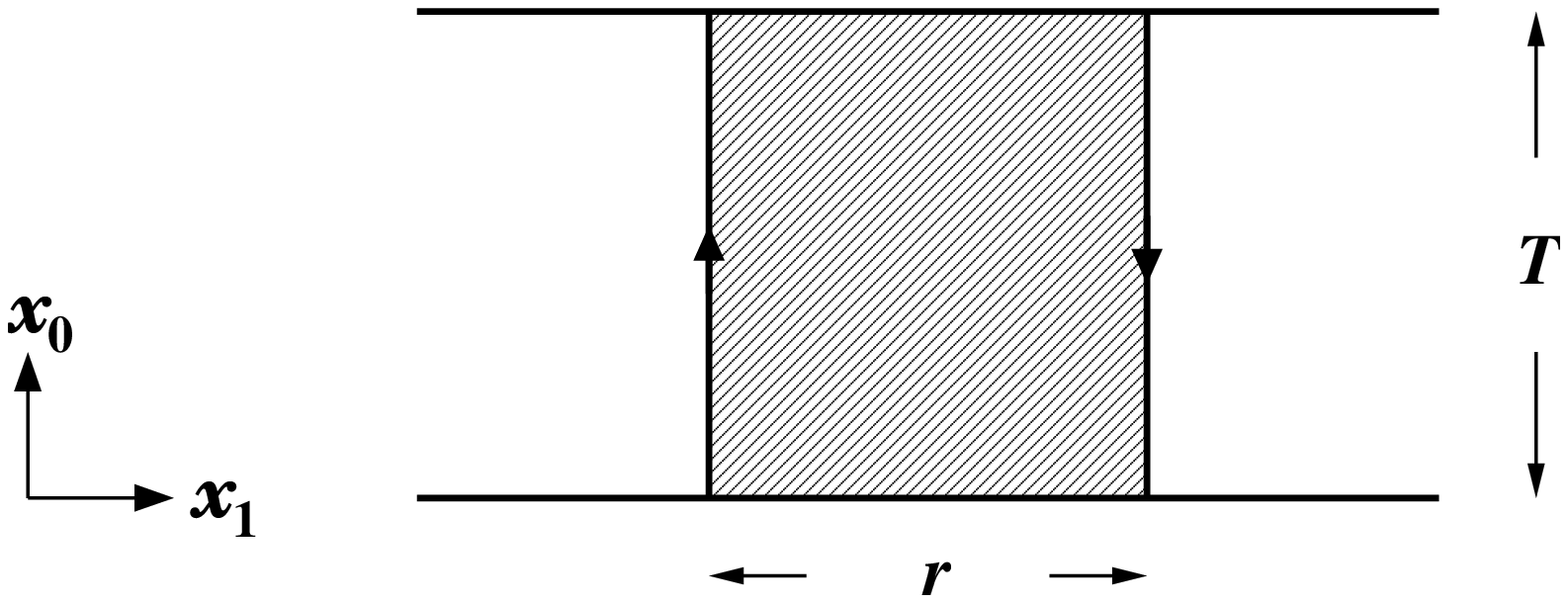}
\vskip0.4cm
\figurecaption{%
In the effective string theory, the 
Polyakov loop correlation function is ob\-tained 
by expanding the associated string partition function about the minimal
surface (shaded area) spanned by the loops. 
}
\vskip0.0cm
}
\endinsert

From a purely technical point of view, the effective string theory
outlined above is a two-dimensional quantum field theory on 
the world-sheet parameter space with specified boundary conditions.
Since the coefficients of the possible interaction terms 
are dimensionful,
the perturbation expansion generated by the effective action 
is non-renormalizable,
and counterterms of increasing dimension
thus need to be included to cancel the divergent
parts of the loop diagrams.

\subsection 3.2 Leading-order approximation

If we neglect the interaction terms in the effective action (3.2),
the functional integral over all vector fields $h(z)$ 
can be evaluated exactly. This
leads to the expression
\equation{
  \left\langle P(x)^{\ast}P(y)\right\rangle=
  \rme^{-\sigma rT-\mu T}
  \left[\det(-\Delta)\right]^{-{1\over2}(d-2)},
  \enum
}
where $\Delta$ denotes the laplacian
on a two-dimensional cylinder of height $r$ and circumference $T$
with Dirichlet boundary conditions.
The mass $\mu$ that appears in this formula is a free parameter of the
string model whose value has no particular physical meaning.
In the gauge theory there is in fact a similar normalization ambiguity,
once the need for renormalization of the Polyakov loop
correlation function is taken into account 
[\ref{DotsenkoVergeles},\ref{BrandtEtAl}].

The determinant of the laplacian can be computed 
explicitly using the $\zeta$-function finite-part prescription
or one of the standard regularization methods.
Such calculations have been performed
in refs.~[\ref{DietzFilk},\ref{AmbjornEtAl},\ref{CaselleEtAl}], for example,
and also in many papers on conformal field theory and 
fundamental string theory
(see refs.~[\ref{Cardy},\ref{Ginsparg}]
for a review and a list of references). 
In terms of the Dedekind $\eta$-function,
\equation{
  \eta(q)=q^{{1\over24}}\prod_{n=1}^{\infty}(1-q^n),
  \enum
}
the result is
\equation{
  \det(-\Delta)=\eta(q)^2,
  \qquad
  q=\rme^{-\pi T/r},
  \enum
}
from which we immediately deduce that 
the string amplitude on the right-hand side of eq.~(3.3) admits
an expansion of the form (2.7), with energies and weights
\equation{
  E_n=\sigma r+\mu+{\pi\over r}\left\{-\frac{1}{24}(d-2)+n\right\},
  \enum\next{2.5ex}
  w_0=1,\quad w_1=d-2,\quad w_2=\frac{1}{2}(d-2)(d+1),\quad\hbox{etc.}
  \enum
}
In particular,
to this order of the effective string theory,
the static quark potential
is given by eq.~(1.1)
(without any corrections proportional to higher powers of $1/r$)
and the energy gap $\Delta E$ is equal to $\pi/r$.

\subsection 3.3 Higher orders \& renormalization

Quantum field theories on space-time manifolds with boundaries 
require, in general, the inclusion of terms in the action that
are localized at the boundary [\ref{QFTbI},\ref{QFTbII}].
In the present case the simplest term of this type is
\equation{
   S_1=\frac{1}{4}b\int_0^T\rmd z_0
   \left\{
   \left(\partial_1h\partial_1h\right)_{z_1=0}+
   \left(\partial_1h\partial_1h\right)_{z_1=r}
   \right\},
   \enum
}
where $b$ is a parameter of dimension $[\hbox{length}]$.
The effect of $S_1$
on the Polyakov loop correlation function can be easily worked out
to first order in $b$ using the $\zeta$-function finite-part 
prescription, for example. For the static potential 
and the energy gap the calculation yields
\equation{
  V(r)=\sigma r+\mu
  -{\pi\over 24r}\left(d-2\right)\left(1+b/r\right),
  \enum\next{2.5ex}
  \Delta E={\pi\over r}\left(1+b/r\right).
  \enum
}
As expected these corrections are negligible compared to the 
leading-order terms at sufficiently large values of $r$.

It can be shown that the boundary action $S_1$ is the 
only term with a coefficient of dimension~$1$
that can occur in the effective action.
There are just a few more terms with coefficients of dimension
$[\hbox{length}]^2$, such as
\equation{
   S_2=\frac{1}{4}c_2\int_0^T\int_0^r\rmd z_0\rmd z_1
   \left(\partial_ah\partial_ah\right)
   \left(\partial_bh\partial_bh\right).
   \enum
}
They give rise to corrections to the static quark
potential proportional to $1/r^3$, i.e.~to terms that 
are even more suppressed at large distances.

A more careful discussion of the higher-order contributions must take
into account the fact that the theory needs to be regularized.
This introduces another scale, the ultra-violet cutoff $\Lambda$,
and corrections proportional to $(c_2/r^3)(\Lambda r)^p$, for example,
must then be expected to arise.
However, as in any other non-renormalizable field theory, it must be 
possible to cancel the divergent contributions by a renormalization of 
$\sigma$, $\mu$ and the coefficients 
of the interaction terms.

For illustration we note that a 
term proportional to $c_2\Lambda/r^2$ actually does occur 
if a lattice regularization is used. 
In this case the inclusion of the 
boundary term $S_1$ is thus required to be able to renormalize
the theory. 
The bottom line is then that equations like (3.9) and (3.10) hold
to all orders of the renormalized effective theory,
up to corrections proportional to higher powers of $1/r$.

\section 4. Numerical evaluation of $V(r)$ and its derivatives

Using the simulation algorithm of ref.~[\ref{MultiLevel}], 
the Polyakov loop
correlation function can be computed at values of $T$ and $r$ that
previously appeared to be inaccessible. In particular, 
when $T$ is increased at fixed $r$, the computer time required
for a spe\-ci\-fied {\it relative}\/ 
statistical error is only growing approximately like $T^3$,
while the signal is exponentially decreasing.
This enables us to obtain the static quark
potential $V(r)$ via eq.~(2.8) with small errors.

The effective string theory predicts that the force $F(r)=V'(r)$
decreases towards the string tension $\sigma$ at large distances $r$
with a rate proportional to $1/r^2$.
In this paper our principal goal is to find out
whether the numerical data are consistent with this and whether
the slope
\equation{
  c(r)=\frac{1}{2}r^3F'(r)
  \enum
}
converges to the value of the universal coefficient $\sc$  
in eq.~(1.1). We are, therefore, more interested in the 
derivatives of the static potential than in the potential itself.

\subsection 4.1 Lattice definition of $F(r)$ and $c(r)$

There is no unique way to ``differentiate" lattice functions,
but a sensible prescription should evidently be such that 
the correct continuum limit is obtained and that no 
artificially large lattice effects are introduced.
The definition of $F(r)$ proposed by Sommer [\ref{SommerScaleA}]
fulfils this condition and is easily extended to $c(r)$. We are thus
led to define
\equation{
  F(\rbar)=\{V(r)-V(r-a)\}/a,
  \enum
  \next{2ex}
  c(\rtilde)=\frac{1}{2}\rtilde^3\{V(r+a)+V(r-a)-2V(r)\}/a^2,
  \enum
}
where $\rbar(r)$ and $\rtilde(r)$ satisfy
\equation{
   \rbar=r-\frac{1}{2}a+\rmO(a^2),
   \enum\next{2ex}
   \rtilde=r+\rmO(a^2).
   \enum
}
The precise choice of these functions takes into account
the approximate shape of the potential, so that 
an enhancement of lattice effects is avoided, particularly at small 
distances [\ref{SommerScaleA},\ref{SilviaRainer}]. Explicitly we set
\equation{
  \rbar^{2-d}=
  2\pi(d-2)\{G(r-a)-G(r)\}/a,
  \enum
  \next{2ex}
  \rtilde^{1-d}=
  2\pi\{G(r+a)+G(r-a)-2G(r)\}/a^2,
  \enum
}
where $G(r)$ denotes the value of the Green function of the
lattice laplacian in $d-1$ dimensions at the point $(r,0,\ldots,0)$
(for a detailed discussion of the Green function
in two and three dimensions,
see refs.~[\ref{Shin},\ref{SilviaRainer}] for example).

\subsection 4.2 Systematic errors

Lattice spacing effects, the omission of the 
contribution of the higher-energy states in eq.~(2.8)
and finite-volume effects are
sources of systematic uncertainties in our calculations.
As usual the lattice effects can be controlled by 
performing simulations at several values of the lattice spacing,
and we shall return to this issue in sect.~5 when we discuss
the simulation results.

To estimate the error $\varepsilon$ in eq.~(2.8), 
some information on the energy values $E_n$ and 
the associated weights $w_n$ is needed.
Accurate data for the lowest excited levels 
were first obtained by Michael and Perantonis
[\ref{MichaelPerantonis}], 
and more precise and extensive studies have later been
reported by Juge, Kuti and Morningstar
[\ref{KutiEtAlI},\ref{KutiEtAlII}]. 
The results show that, in four dimensions, the gap $\Delta E$
is about $2.2/r$ at the Sommer reference scale $r=r_0$.
It then approaches $\pi/r$ around $r=3r_0$ and 
stays above this value at still larger distances.
The associated weight $w_1$ is $2$ as in the effective string theory.
Using this and the fact that the higher levels
are well separated from $E_1$, the error
in eq.~(2.8) can be estimated straightforwardly.

There are unfortunately no results on the spectrum
of the excited levels in three dimensions, and we can only 
refer to the analogous case of the SU(2) theory, where
the situation does not appear to be very different
from the one described above
[\ref{KutiEtAlII}].
We have, therefore, estimated the error in eq.~(2.8)
in the same way, with $w_1=1$,
and checked the estimate by 
performing simulations at various values of $T$.

Finite-volume effects can arise
when the colour flux tube is squeezed
in the transverse directions by the finite extent of the lattice.
In all our calculations the spatial lattice size $L$ was at least
$20$ times larger
than the bulk correlation length [\ref{Michael},\ref{Teper}],
and these effects were thus suppressed by many
orders of magnitude.
Another systematic effect can be traced back to the 
presence of states where
the flux tube is winding ``around the world". 
The energy of these states is at least $V(L-r)$, and
their contribution to the error
in eq.~(2.8) will therefore be small if $L$ and $r$ are such
that $V(L-r)-V(r)$ is large.

\subsection 4.3 Statistical errors

The parameters of the simulation algorithm of ref.~[\ref{MultiLevel}]
can be chosen so that subse\-quent ``measurements" of the Polyakov loop
correlation function may be assumed to be statistically independent.
We have done so and later confirmed the independence of the data
by determining the integrated autocorrelation time.
The standard single-elimination jackknife analysis was then applied to 
calculate the statistical errors.

An important property of the data series generated in this way is
that the values of the static potential at 
different distances
turn out to be very strongly correlated. As a consequence
there is a significant cancellation of statistical errors
when the force $F(r)$ and the slope $c(r)$ are calculated.
It is then also profitable to keep track of the statistical correlations
in the further analysis of the data.

\topinsert
\newdimen\digitwidth
\setbox0=\hbox{\rm 0}
\digitwidth=\wd0
\catcode`@=\active
\def@{\kern\digitwidth}
\tablecaption{Simulation parameters}
\vskip1.0ex
$$\vbox{\settabs\+&%
                  xxxxx&&
                  xxxxx&&
                  xxxxxxx&&
                  xxxxxxx&&
                  xxxxxxxx&&
                  xxxxxx&&
                  xxxxxx&
                  x&\cr
\thicktablerule
\vskip1ex
                \+& \hfill $d$ \hfill
                 && \hfill $\beta$\hfill
                 && \hfill $r_0/a$\hfill
                 && \hfill $a$ [fm]\hfill
                 && \hfill $r_{\rm max}/a$\hfill
                 && \hfill $T/a$\hfill
                 && \hfill $L/a$\hfill
                 && \cr
\vskip1.0ex
\thintablerule
\vskip1.5ex
  \+& \hfill $4$\hfill
  &&  \hfill $5.70$\hfill 
  &&  \hfill $2.93$\hfill
  &&  \hfill $0.171$\hfill
  &&  \hfill $7$\hfill 
  &&  \hfill $24$\hfill
  &&  \hfill $18$\hfill 
  &\cr
\vskip0.3ex
  \+& \hfill $4$\hfill
  &&  \hfill $5.85$\hfill 
  &&  \hfill $4.09$\hfill
  &&  \hfill $0.122$\hfill
  &&  \hfill $9$\hfill 
  &&  \hfill $36$\hfill
  &&  \hfill $24$\hfill 
  &\cr
\vskip0.3ex
  \+& \hfill $4$\hfill
  &&  \hfill $6.00$\hfill 
  &&  \hfill $5.39$\hfill
  &&  \hfill $0.093$\hfill
  &&  \hfill $12$\hfill 
  &&  \hfill $48$\hfill
  &&  \hfill $30$\hfill 
  &\cr
\vskip0.3ex
  \+& \hfill $3$\hfill
  &&  \hfill $11.0$\hfill 
  &&  \hfill $3.30$\hfill
  &&  \hfill $0.152$\hfill
  &&  \hfill $9$\hfill 
  &&  \hfill $32$\hfill
  &&  \hfill $24$\hfill 
  &\cr
\vskip0.3ex
  \+& \hfill $3$\hfill
  &&  \hfill $15.0$\hfill 
  &&  \hfill $4.83$\hfill
  &&  \hfill $0.104$\hfill
  &&  \hfill $12$\hfill 
  &&  \hfill $48$\hfill
  &&  \hfill $32$\hfill 
  &\cr
\vskip0.3ex
  \+& \hfill $3$\hfill
  &&  \hfill $20.0$\hfill 
  &&  \hfill $6.71$\hfill
  &&  \hfill $0.075$\hfill
  &&  \hfill $14$\hfill 
  &&  \hfill $60$\hfill
  &&  \hfill $36$\hfill 
  &\cr
\vskip1.5ex
\thicktablerule
}$$
\endinsert

\section 5. Simulation results

In table~1 we list the values of the coupling $\beta=6a^{d-4}/g_0^2$
and other important para\-me\-ters of the simulations that we  
have performed. The entries in the last three columns are the 
maximal distance and the associated lattice sizes 
where we have accurate results for the Polyakov loop 
correlation function.
At smaller distances, the lattices do not need to be as large,
but were always chosen so
that the systematic errors (other than the lattice effects) 
are sufficiently suppressed.
Further algorithmic details 
and a collection of data tables are included in appendices A and B.

In the following we first discuss the results 
at the smallest values of the lattice spacing
and shall then argue (in subsect.~5.3) that 
the lattice effects at these points are already very small.

\topinsert
\vbox{
\vskip0.0cm
\epsfxsize=9.5cm\hskip0.9cm\epsfbox{figure2.eps}
\vskip0.3cm
\figurecaption{%
Force $F(r)$ in the four-dimensional theory versus $1/r^2$ at $\beta=6.0$.  
Physical units are set by the Sommer scale $r_0=0.5$ fm [\ref{SommerScaleA}]
and the dotted line is a linear fit to the four points at the largest
distances. Errors are smaller than the data symbols.
}
\vskip0.0cm
}
\endinsert

\subsection 5.1 $F(r)$ and $c(r)$ in four dimensions

One of the striking results of our computations is that 
the force $F(r)$ appears to be a linear 
function of $1/r^2$ in the range $r\geq0.5$~fm
(see fig.~2). 
The errors on the data points shown in this figure increase
from below $0.04\%$ to about $0.12\%$ at the left-most point,
and the four points at the  
largest values of $r$ lie on a straight line 
to this level of accuracy.

There is still some curvature in the data,
which is more clearly seen in fig.~3 where we plot the slope $c(r)$.
The errors on these points are at most $2.5\%$ and significantly
less than this at the smaller distances.
Note that $c(r)$ can be interpreted as a running coupling,
\equation{
  c(r)=-\frac{4}{3}\alpha(1/r),
  \qquad \alpha(\mu)={\gbar(\mu)^2\over 4\pi},
  \enum
}
that satisfies the usual perturbative renormalization group equation
\equation{
  \mu{\partial \gbar\over\partial\mu}=-\sum_{n=0}^{\infty}
  b_n\gbar^{2n+3}.
  \enum
}
The coefficients $b_n$ are known up to three-loop order 
in this scheme [\ref{Fischler}--\ref{Melles}],
and using these we can integrate eq.~(5.2) from $r=0$,
taking the $\Lambda$ parameter determined in 
refs.~[\ref{AlphaI},\ref{AlphaII}] as the initial condition. 

In the related case of the force $F(r)$,
the range of applicability of the perturbation expansion 
extends to distances of around $0.2$ or at most $0.3$~fm
[\ref{SilviaRainerII}].
We refer to this paper for further details and 
only remark that the error on the $\Lambda$~para\-me\-ter alone
results in a significant spread of the three-loop curves
shown in fig.~3 (shaded area on the left).
It is in any case evident from the plot
that $c(r)$ breaks away
from the perturbative behaviour near $r=0.2$~fm.

\topinsert
\vbox{
\vskip0.0cm
\epsfxsize=9.0cm\hskip1.15cm\epsfbox{figure3.eps}
\vskip0.3cm
\figurecaption{%
Slope $c(r)$ in three and four dimensions at $\beta=20.0$ 
and $\beta=6.0$ respectively. The curves on the left are 
obtained from perturbation theory, while those on the right
derive from the string theory formula (3.9) with $b=0$
and $b=0.04$~fm.
}
\vskip0.0cm
}
\endinsert

At larger distances the data show a shallow minimum and then 
gently increase to about $-0.293$ at $r=1$~fm.
This is still $12\%$ away from 
$\sc=-0.262$, but not really inconsistent
with the effective string theory, because the correction proportional to $b$
in eq.~(3.9) (and possibly higher-order terms) may not be negligible
at these distances. As shown in fig.~3 we can in fact easily make contact
with the data if we allow for such a correction.

\subsection 5.2 Three-dimensional theory

In three dimensions 
the static potential at short distances 
has a completely different shape,
but the effective string theory
predicts the same behaviour at large distances, with
a coefficient $\sc$ that is precisely half as big as in four dimensions.

Our numerical results fully agree
with this (see fig.~3). 
No $b/r$ correction is required in this case
to match the data points at the largest distances, while
at the smaller values of $r$ we observe a smooth transition
from perturbative to string behaviour.
In this theory the perturbation 
expansion to two-loop order reads [\ref{Schroder}]
\equation{
  c(r)=-{1\over3\pi}g^2r+\rmO(g^6),
  \enum
}
where $g$ denotes the gauge coupling in the continuum limit.
To express the coupling in physical units, we used the 
conversion factor $g^2r_0=2.2$ (which can be inferred from our data) 
and set $r_0=0.5$~fm as usual.

\subsection 5.3 Lattice effects

In both three and four dimensions, 
we have simulated lattices at different values of 
the lattice spacing in order to study the approach of
the force $F(r)$ and the slope $c(r)$
to the continuum limit (see table~1). 
The comparison of the data produced in these simulations
requires all dimensionful quantities to be scaled 
to some physical units. 
It is natural to take the Sommer radius $r_0$ as the basic 
reference scale
in the present context, and we shall thus be interested in determining
the lattice-spacing dependence of the dimensionless functions
$r_0^2F(xr_0)$ and $c(xr_0)$ in a range of the 
scale factor $x$.

Since the distances $xr_0$ are in general fractional multiples
of the lattice spacing, an interpolation formula must be specified.
Given the approximate shape of $F(r)$ and $c(r)$,
we decided to use a three-point polynomial interpolation
in $1/x^2$ and $1/x$ respectively.
Note that it would be logically incorrect to assign a
systematic error to the interpolation, because 
the latter merely defines the functions between 
the lattice points. On the other hand, a poor choice of the 
interpolation formula will lead to artificially enhanced
lattice effects. 

From the tables in appendix B it is clear that the lattice effects 
are quite small at the lattice spacings considered.
In both three and four dimensions,
the residual effects on the lattices with the smallest spacing 
are probably no more than 
about $0.1\%$ in the case of the force $r_0^2F(xr_0)$ 
and $1-2\%$ in the case of the slope $c(xr_0)$.
The theoretically expected scaling proportional to $a^2$ 
[\ref{Symanzik},\ref{SilviaRainer}]
is, however, not observed, at least not in four dimensions, where the 
lattice effects at $\beta=5.7$ appear to be larger than
what would be inferred from the other lattices.

For this reason, and since we have data at
only three values of the lattice spacing, we do not attempt to 
perform an extrapolation to the continuum limit. 
We would in any case not expect the extrapolation to have any
impact on the outcome of our work, because the lattice effects
are small and because in four dimensions 
the values of $c(xr_0)$ at the
larger distances $xr_0$ would only be moved closer to $\sc$.

\section 6. Conclusions

The results reported in this paper
show that string behaviour 
in the static potential sets in
at quark-antiquark separations around $0.5$~fm.
In three dimensions the confirmation of the 
string theory formula (1.1)
is particularly impressive, while in four dimensions
a small higher-order correction needs to be included to 
match the data points at the largest distances in fig.~3.

The observed agreement with the effective string theory
is somewhat surprising and in fact difficult to understand,
because the physical picture of a thin fluctuating flux tube is
hardly correct at these distances.
Perhaps this is an indication of the 
existence of an exact dual formulation of SU($N$)
gauge theories in terms of a fundamental string theory,
along the lines of ref.~[\ref{Polyakov}] for example,
from which the effective theory derives.
Note that not only the general form of the effective action,
but also the values of the coefficients of the higher-dimensional terms 
would be predictable in this case (to the extent the fundamental
string theory is tractable).

Another logical possibility is that 
eq.~(1.1) may have a different origin and that 
the proper interpretation of our results
still needs to be found.
It has been pointed out in this connection
that the spectrum of the excited levels $E_n$
does not appear to agree with the one obtained
at leading order of the effective string theory
[\ref{KutiEtAlI},\ref{KutiEtAlII}].
Whether this rules out
the effective theory is not completely obvious, however,
because the observed discrepancies (which are significant
even at distances~$r\geq2$~fm) could be
a result of the higher-order corrections.
In particular, the accidental degeneracies of the spectrum
at leading order probably disappear once the non-linear
interactions are switched on.

A systematic study of the higher-order effects 
would now be required to resolve this issue.
Accurate calculations of the lowest
energy values in each symmetry sector may then
already provide enough information to determine 
the coefficients of the few leading terms in the effective action.
It would evidently also be interesting to see whether our results
carry over to other non-abelian gauge groups and to all representations
of the Wilson lines with non-trivial transformation behaviour
under the centre of the gauge group.
We finally note that
the computation of $c(r)$ at 
distances larger than reported here rapidly requires enormous amounts
of computer time, 
because the significance loss in eq.~(4.3) (which grows proportionally
to $\sigma r^4/a^2$)
must be compensated by increasing the statistics.

\vskip1ex\noindent
{\it Acknowledgements:}\/
We are indebted to 
Christina Diamantini, Julius Kuti, Rainer Sommer, Poul Olesen 
and Yaron Oz for helpful correspondence and discussions,
and to Julius in particular for sending us a draft of his
forthcoming long paper with Jimmy Juge and Colin Morningstar
on the spectrum of the energy values~$E_n$.
Our computations have been performed on PC-clusters at 
DESY (Hamburg and Zeuthen), at the Max-Planck-Institut
f\"ur Physik in 
Munich and at the Fermi Institute in Rome. 
We wish to thank the directors of these institutions
for having made this possible,
and Peter Breitenlohner, Filippo Palombi, Peter Wegner and Hartmut Wittig
for technical support.

\appendix A. Simulation algorithm

The multilevel algorithm that was used in the present study
is described in detail in ref.~[\ref{MultiLevel}].
Here we merely wish to add a few comments
on performance and opti\-mi\-zation issues.

All tests of the algorithm 
reported in ref.~[\ref{MultiLevel}] were carried out
on four-dimensional lattices at $\beta=5.7$.
We now confirm that it performs well
also at larger values of $\beta$ and in three dimensions. 
For illustration we quote 
the computation of the Polyakov loop correlation function
at $\beta=6.0$, $T=48a$ and $r=12a$.
This is an extreme case where the area of the minimal surface bounded 
by the loops reaches $5\,\fm^2$.
The correlation function is very small at this point
(about $1.1\times10^{-25}$), but with
one month of running time on a PC-cluster with 24 Pentium~4 processors
it was possible to determine its value to an accuracy 
better than $3\%$.

Contrary to what may be understood from ref.~[\ref{MultiLevel}],
the more complicated forms of the algorithm, with 
many levels of nested averages, are often not worth while.
Most of our results have in fact
been obtained with only one level,
but a relatively large number (up to a few thousands)
of ``measurements" of the two-link operators
on the associated time slices.
An important side effect of such high numbers of time-slice updates is 
that the statistical correlations between the 
data at the distances covered in a given run
tend to be enhanced (cf.~subsect.~4.3).

For future applications of the multilevel algorithm 
at larger distances $r$, it may
be necessary to study and improve its efficiency at
the level of the time-slice averages.
There is currently no understanding of the autocorrelation times
in this subsystem, and 
rather than increasing the statistics, there may be better 
ways to cope with the
exponential decay of the averaged two-link operators.

\appendix B. Data tables

In tables 2--7 we only list the most important results
for the force $F(r)$ and the slope $c(r)$. 
More detailed data tables, including
error correlation matrices, may be obtained from the authors
in electronic form.

\vskip0.3cm

\vbox{
\newdimen\digitwidth
\setbox0=\hbox{\rm 0}
\digitwidth=\wd0
\catcode`@=\active
\def@{\kern\digitwidth}
\tablecaption{Simulation results at $\beta=6.0$ in $4$ dimensions}
\vskip1.0ex
$$\vbox{\settabs\+&%
                  xxxxxx&&
                  xxxxxxxx&&
                  xxxxxxxxxxxxxx&&
                  xxxxxxxx&&
                  xxxxxxxxxxxxx&&
                  x&\cr
\thicktablerule
\vskip1ex
                \+& \hfill $r/a$ \hfill
                 && \hfill $\hskip0.8ex\rbar/a$\hfill
                 && \hfill $a^2F(\rbar)$\hfill
                 && \hfill $\rtilde/a$\hfill
                 && \hfill $\hskip-0.8ex c(\rtilde)$\hfill
                 && \cr
\vskip1.0ex
\thintablerule
\vskip1.5ex
  \+& \hfill $3$\hfill
  &&  \hfill $@2.277$\hfill 
  &&  \hfill $0.102491(20)$\hfill
  &&  \hfill $2.700$\hfill
  &&  \hfill $0.28184(10)$\hfill
  &\cr
\vskip0.3ex
  \+& \hfill $4$\hfill
  &&  \hfill $@3.312$\hfill 
  &&  \hfill $0.073854(22)$\hfill
  &&  \hfill $3.729$\hfill
  &&  \hfill $0.29724(18)$\hfill
  &\cr
\vskip0.3ex
  \+& \hfill $5$\hfill
  &&  \hfill $@4.359$\hfill 
  &&  \hfill $0.062399(24)$\hfill
  &&  \hfill $4.786$\hfill
  &&  \hfill $0.30339(38)$\hfill
  &\cr
\vskip0.3ex
  \+& \hfill $6$\hfill
  &&  \hfill $@5.393$\hfill 
  &&  \hfill $0.056871(25)$\hfill
  &&  \hfill $5.833$\hfill
  &&  \hfill $0.30515(70)$\hfill
  &\cr
\vskip0.3ex
  \+& \hfill $7$\hfill
  &&  \hfill $@6.414$\hfill 
  &&  \hfill $0.053800(27)$\hfill
  &&  \hfill $6.864$\hfill
  &&  \hfill $0.3041(12)@$\hfill
  &\cr
\vskip0.3ex
  \+& \hfill $8$\hfill
  &&  \hfill $@7.428$\hfill 
  &&  \hfill $0.051924(30)$\hfill
  &&  \hfill $7.886$\hfill
  &&  \hfill $0.3033(17)@$\hfill
  &\cr
\vskip0.3ex
  \+& \hfill $9$\hfill
  &&  \hfill $@8.438$\hfill 
  &&  \hfill $0.050687(34)$\hfill
  &&  \hfill $8.901$\hfill
  &&  \hfill $0.3008(28)@$\hfill
  &\cr
\vskip0.3ex
  \+& \hfill $10$\hfill
  &&  \hfill $@9.445$\hfill 
  &&  \hfill $0.049834(38)$\hfill
  &&  \hfill $9.912$\hfill
  &&  \hfill $0.2992(80)@$\hfill
  &\cr
\vskip0.3ex
  \+& \hfill $11$\hfill
  &&  \hfill $10.451$\hfill 
  &&  \hfill $0.049213(46)$\hfill
  &&  \hfill $$\hfill
  &&  \hfill $$\hfill
  &\cr
\vskip0.3ex
  \+& \hfill $12$\hfill
  &&  \hfill $11.455$\hfill 
  &&  \hfill $0.048839(72)$\hfill
  &&  \hfill $$\hfill
  &&  \hfill $$\hfill
  &\cr
\vskip1.0ex
\thicktablerule
}$$
}

\vfill

\topinsert
\newdimen\digitwidth
\setbox0=\hbox{\rm 0}
\digitwidth=\wd0
\catcode`@=\active
\def@{\kern\digitwidth}
\tablecaption{Simulation results at $\beta=20.0$ in $3$ dimensions}
\vskip1.0ex
$$\vbox{\settabs\+&%
                  xxxxxx&&
                  xxxxxxxx&&
                  xxxxxxxxxxxxxx&x&
                  xxxxxxxx&&
                  xxxxxxxxxxxxxx&&
                  x&\cr
\thicktablerule
\vskip1ex
                \+& \hfill $r/a$ \hfill
                 && \hfill $\hskip0.8ex\rbar/a$\hfill
                 && \hfill $a^2F(\rbar)$\hfill
                 && \hfill $\hskip1ex\rtilde/a$\hfill
                 && \hfill $c(\rtilde)\hskip1ex$\hfill
                 && \cr
\vskip1.0ex
\thintablerule
\vskip1.5ex
  \+& \hfill $3$\hfill
  &&  \hfill $@2.379$\hfill 
  &&  \hfill $0.0508627(32)$\hfill
  &&  \hfill $@2.808$\hfill
  &&  \hfill $0.081536(41)$\hfill
  &\cr
\vskip0.3ex
  \+& \hfill $4$\hfill
  &&  \hfill $@3.407$\hfill 
  &&  \hfill $0.0434942(32)$\hfill
  &&  \hfill $@3.838$\hfill
  &&  \hfill $0.099772(77)$\hfill
  &\cr
\vskip0.3ex
  \+& \hfill $5$\hfill
  &&  \hfill $@4.432$\hfill 
  &&  \hfill $0.0399646(33)$\hfill
  &&  \hfill $@4.875$\hfill
  &&  \hfill $0.112270(96)$\hfill
  &\cr
\vskip0.3ex
  \+& \hfill $6$\hfill
  &&  \hfill $@5.448$\hfill 
  &&  \hfill $0.0380274(35)$\hfill
  &&  \hfill $@5.902$\hfill
  &&  \hfill $0.12057(13)@$\hfill
  &\cr
\vskip0.3ex
  \+& \hfill $7$\hfill
  &&  \hfill $@6.458$\hfill 
  &&  \hfill $0.0368546(38)$\hfill
  &&  \hfill $@6.920$\hfill
  &&  \hfill $0.12596(23)@$\hfill
  &\cr
\vskip0.3ex
  \+& \hfill $8$\hfill
  &&  \hfill $@7.464$\hfill 
  &&  \hfill $0.0360941(41)$\hfill
  &&  \hfill $@7.932$\hfill
  &&  \hfill $0.12945(41)@$\hfill
  &\cr
\vskip0.3ex
  \+& \hfill $9$\hfill
  &&  \hfill $@8.469$\hfill 
  &&  \hfill $0.0355749(45)$\hfill
  &&  \hfill $@8.941$\hfill
  &&  \hfill $0.13100(55)@$\hfill
  &\cr
\vskip0.3ex
  \+& \hfill $10$\hfill
  &&  \hfill $@9.473$\hfill 
  &&  \hfill $0.0352080(49)$\hfill
  &&  \hfill $@9.948$\hfill
  &&  \hfill $0.13184(69)@$\hfill
  &\cr
\vskip0.3ex
  \+& \hfill $11$\hfill
  &&  \hfill $10.475$\hfill 
  &&  \hfill $0.0349401(55)$\hfill
  &&  \hfill $10.953$\hfill
  &&  \hfill $0.13276(92)@$\hfill
  &\cr
\vskip0.3ex
  \+& \hfill $12$\hfill
  &&  \hfill $11.478$\hfill 
  &&  \hfill $0.0347382(62)$\hfill
  &&  \hfill $11.957$\hfill
  &&  \hfill $0.1321(17)@@$\hfill
  &\cr
\vskip0.3ex
  \+& \hfill $13$\hfill
  &&  \hfill $12.480$\hfill 
  &&  \hfill $0.0345838(73)$\hfill
  &&  \hfill $12.961$\hfill
  &&  \hfill $0.1271(34)@@$\hfill
  &\cr
\vskip0.3ex
  \+& \hfill $14$\hfill
  &&  \hfill $13.481$\hfill 
  &&  \hfill $0.0344673(92)$\hfill
  &&  \hfill $$\hfill
  &&  \hfill $$\hfill
  &\cr
\vskip1.0ex
\thicktablerule
}$$
\endinsert

\vbox{}
\vfill\eject

\topinsert
\newdimen\digitwidth
\setbox0=\hbox{\rm 0}
\digitwidth=\wd0
\catcode`@=\active
\def@{\kern\digitwidth}
\tablecaption{Values of $r_0^2F(xr_0)$ in $4$ dimensions}
\vskip1.0ex
$$\vbox{\settabs\+&%
                  xxxxxxx&&
                  xxxxxxxxxxxxx&&
                  xxxxxxxxxxxxx&&
                  xxxxxxxxxxxxx&&
                  x&\cr
\thicktablerule
\vskip1ex
                \+& \hfill $x$ \hfill
                 && \hfill $\beta=5.70$\hfill
                 && \hfill $\beta=5.85$\hfill
                 && \hfill $\beta=6.00$\hfill
                 && \cr
\vskip1.0ex
\thintablerule
\vskip1.5ex
  \+& \hfill $0.8$\hfill
  &&  \hfill $1.82181(23)$\hfill 
  &&  \hfill $1.82203(57)$\hfill
  &&  \hfill $1.82041(16)$\hfill
  &\cr
\vskip0.3ex
  \+& \hfill $1.0$\hfill
  &&  \hfill $1.65@@@@@@\hskip0.6ex$\hfill 
  &&  \hfill $1.65@@@@@@\hskip0.6ex$\hfill
  &&  \hfill $1.65@@@@@@\hskip0.6ex$\hfill
  &\cr
\vskip0.3ex
  \+& \hfill $1.2$\hfill
  &&  \hfill $1.55341(12)$\hfill 
  &&  \hfill $1.55602(34)$\hfill
  &&  \hfill $1.55691(18)$\hfill
  &\cr
\vskip0.3ex
  \+& \hfill $1.4$\hfill
  &&  \hfill $1.49372(23)$\hfill 
  &&  \hfill $1.49983(49)$\hfill
  &&  \hfill $1.50100(36)$\hfill
  &\cr
\vskip0.3ex
  \+& \hfill $1.6$\hfill
  &&  \hfill $1.45561(31)$\hfill 
  &&  \hfill $1.46369(60)$\hfill
  &&  \hfill $1.46476(51)$\hfill
  &\cr
\vskip0.3ex
  \+& \hfill $1.8$\hfill
  &&  \hfill $1.42963(39)$\hfill 
  &&  \hfill $1.43920(76)$\hfill
  &&  \hfill $1.44010(70)$\hfill
  &\cr
\vskip0.3ex
  \+& \hfill $2.0$\hfill
  &&  \hfill $1.41122(47)$\hfill 
  &&  \hfill $1.4217(10)@$\hfill
  &&  \hfill $1.4230(11)@$\hfill
  &\cr
\vskip1.0ex
\thicktablerule
}$$
\endinsert

\topinsert
\newdimen\digitwidth
\setbox0=\hbox{\rm 0}
\digitwidth=\wd0
\catcode`@=\active
\def@{\kern\digitwidth}
\tablecaption{Values of $r_0^2F(xr_0)$ in $3$ dimensions}
\vskip1.0ex
$$\vbox{\settabs\+&%
                  xxxxxxx&&
                  xxxxxxxxxxxxxx&&
                  xxxxxxxxxxxxxx&&
                  xxxxxxxxxxxxxx&&
                  x&\cr
\thicktablerule
\vskip1ex
                \+& \hfill $x$ \hfill
                 && \hfill $\beta=11.0$\hfill
                 && \hfill $\beta=15.0$\hfill
                 && \hfill $\beta=20.0$\hfill
                 && \cr
\vskip1.0ex
\thintablerule
\vskip1.5ex
  \+& \hfill $0.8$\hfill
  &&  \hfill $1.720701(26)$\hfill 
  &&  \hfill $1.719410(57)$\hfill
  &&  \hfill $1.718083(52)$\hfill
  &\cr
\vskip0.3ex
  \+& \hfill $1.0$\hfill
  &&  \hfill $1.65@@@@@@@\hskip0.6ex$\hfill 
  &&  \hfill $1.65@@@@@@@\hskip0.6ex$\hfill
  &&  \hfill $1.65@@@@@@@\hskip0.6ex$\hfill
  &\cr
\vskip0.3ex
  \+& \hfill $1.2$\hfill
  &&  \hfill $1.608316(37)$\hfill 
  &&  \hfill $1.610380(53)$\hfill
  &&  \hfill $1.610988(73)$\hfill
  &\cr
\vskip0.3ex
  \+& \hfill $1.4$\hfill
  &&  \hfill $1.582508(55)$\hfill 
  &&  \hfill $1.58583(10)@$\hfill
  &&  \hfill $1.58689(12)@$\hfill
  &\cr
\vskip0.3ex
  \+& \hfill $1.6$\hfill
  &&  \hfill $1.565497(77)$\hfill 
  &&  \hfill $1.56965(14)@$\hfill
  &&  \hfill $1.57112(16)@$\hfill
  &\cr
\vskip0.3ex
  \+& \hfill $1.8$\hfill
  &&  \hfill $1.553860(90)$\hfill 
  &&  \hfill $1.55852(19)@$\hfill
  &&  \hfill $1.56022(22)@$\hfill
  &\cr
\vskip0.3ex
  \+& \hfill $2.0$\hfill
  &&  \hfill $1.54555(11)@$\hfill 
  &&  \hfill $1.55048(24)@$\hfill
  &&  \hfill $1.55271(34)@$\hfill
  &\cr
\vskip1.0ex
\thicktablerule
}$$
\endinsert

\vbox{}
\vfill\eject

\topinsert
\newdimen\digitwidth
\setbox0=\hbox{\rm 0}
\digitwidth=\wd0
\catcode`@=\active
\def@{\kern\digitwidth}
\tablecaption{Values of $c(xr_0)$ in $4$ dimensions}
\vskip1.0ex
$$\vbox{\settabs\+&%
                  xxxxxxx&&
                  xxxxxxxxxxxxxx&&
                  xxxxxxxxxxxxxx&&
                  xxxxxxxxxxxxxx&&
                  x&\cr
\thicktablerule
\vskip1ex
                \+& \hfill $x$ \hfill
                 && \hfill $\beta=5.70$\hfill
                 && \hfill $\beta=5.85$\hfill
                 && \hfill $\beta=6.00$\hfill
                 && \cr
\vskip1.0ex
\thintablerule
\vskip1.5ex
  \+& \hfill $1.0$\hfill
  &&  \hfill $-0.31427(43)$\hfill 
  &&  \hfill $-0.3067(10)@$\hfill
  &&  \hfill $-0.30493(58)$\hfill
  &\cr
\vskip0.3ex
  \+& \hfill $1.2$\hfill
  &&  \hfill $-0.32150(57)$\hfill 
  &&  \hfill $-0.3048(14)@$\hfill
  &&  \hfill $-0.3045(11)@$\hfill
  &\cr
\vskip0.3ex
  \+& \hfill $1.4$\hfill
  &&  \hfill $-0.32113(72)$\hfill 
  &&  \hfill $-0.3036(13)@$\hfill
  &&  \hfill $-0.3039(15)@$\hfill
  &\cr
\vskip0.3ex
  \+& \hfill $1.6$\hfill
  &&  \hfill $-0.3176(10)@$\hfill 
  &&  \hfill $-0.3003(22)@$\hfill
  &&  \hfill $-0.3014(24)@$\hfill
  &\cr
\vskip0.3ex
  \+& \hfill $1.8$\hfill
  &&  \hfill $-0.3150(19)@$\hfill 
  &&  \hfill $-0.2990(50)@$\hfill
  &&  \hfill $-0.2995(64)@$\hfill
  &\cr
\vskip1.0ex
\thicktablerule
}$$
\endinsert

\topinsert
\newdimen\digitwidth
\setbox0=\hbox{\rm 0}
\digitwidth=\wd0
\catcode`@=\active
\def@{\kern\digitwidth}
\tablecaption{Values of $c(xr_0)$ in $3$ dimensions}
\vskip1.0ex
$$\vbox{\settabs\+&%
                  xxxxxxx&&
                  xxxxxxxxxxxxxx&&
                  xxxxxxxxxxxxxx&&
                  xxxxxxxxxxxxxx&&
                  x&\cr
\thicktablerule
\vskip1ex
                \+& \hfill $x$ \hfill
                 && \hfill $\beta=11.0$\hfill
                 && \hfill $\beta=15.0$\hfill
                 && \hfill $\beta=20.0$\hfill
                 && \cr
\vskip1.0ex
\thintablerule
\vskip1.5ex
  \+& \hfill $1.0$\hfill
  &&  \hfill $-0.13225(14)$\hfill 
  &&  \hfill $-0.12696(18)$\hfill
  &&  \hfill $-0.12505(21)$\hfill
  &\cr
\vskip0.3ex
  \+& \hfill $1.2$\hfill
  &&  \hfill $-0.13842(16)$\hfill 
  &&  \hfill $-0.13157(32)$\hfill
  &&  \hfill $-0.12972(40)$\hfill
  &\cr
\vskip0.3ex
  \+& \hfill $1.4$\hfill
  &&  \hfill $-0.14106(30)$\hfill 
  &&  \hfill $-0.13475(40)$\hfill
  &&  \hfill $-0.13144(55)$\hfill
  &\cr
\vskip0.3ex
  \+& \hfill $1.6$\hfill
  &&  \hfill $-0.14153(33)$\hfill 
  &&  \hfill $-0.13552(67)$\hfill
  &&  \hfill $-0.13273(84)$\hfill
  &\cr
\vskip0.3ex
  \+& \hfill $1.8$\hfill
  &&  \hfill $-0.14106(52)$\hfill 
  &&  \hfill $-0.13654(97)$\hfill
  &&  \hfill $-0.1316(18)@$\hfill
  &\cr
\vskip1.0ex
\thicktablerule
}$$
\endinsert

\vbox{}
\vfill \eject

\beginbibliography


\bibitem{Nambu}
Y. Nambu,
Phys. Lett. B80 (1979) 372

\bibitem{WKB}
M. L\"uscher, K. Symanzik, P. Weisz,
Nucl. Phys. B173 (1980) 365

\bibitem{UniversalTerm}
M. L\"uscher,
Nucl. Phys. B180 (1981) 317


\bibitem{SilviaRainer}
S. Necco, R. Sommer,
Nucl. Phys. B622 (2002) 328


\bibitem{CaselleEtAl}
M. Caselle, R. Fiore, F. Gliozzi, M. Hasenbusch, P. Provero,
Nucl. Phys. B486 (1997) 245


\bibitem{MultiLevel}
M. L\"uscher, P. Weisz,
J. High Energy Phys. 09 (2001) 010


\bibitem{AmbjornEtAl}
J. Ambj{\o}rn, P. Olesen, C. Peterson,
Phys. Lett. B142 (1984) 410; Nucl. Phys. B244 (1984) 262

\bibitem{deForcrandEtAl}
P. de Forcrand, G. Schierholz,
H. Schneider, M. Teper,
Phys. Lett. B160 (1985) 137

\bibitem{LuciniTeper}
B. Lucini, M. Teper,
Phys. Rev. D64 (2001) 105019


\bibitem{Wilson}
K. G. Wilson, Phys. Rev. D10 (1974) 2445


\bibitem{TransferI}
M. L\"uscher,
Commun. Math. Phys. 54 (1977) 283

\bibitem{TransferII}
K. Osterwalder, E. Seiler,
Ann. Phys. (NY) 110 (1978) 440

\bibitem{TransferIII}
E. Seiler,
Gauge theories as a problem of constructive quantum field
theory and statistical mechanics, Lecture Notes in Physics 159
(Springer, Berlin, 1982)

\bibitem{TransferIV}
M. L\"uscher,
Selected topics in lattice field theory,
Lectures given at Les Houches (1988),
in: Fields, strings and
critical phenomena, eds. E. Br\'ezin, J. Zinn-Justin
(North-Holland, Amsterdam, 1989)


\bibitem{DotsenkoVergeles}
V. S. Dotsenko, S. N. Vergeles,
Nucl. Phys. B169 (1980) 527

\bibitem{BrandtEtAl}
R. A. Brandt, F. Neri, M. Sato,
Phys. Rev. D24 (1981) 879


\bibitem{DietzFilk}
K. Dietz, T. Filk,
Phys. Rev. D27 (1983) 2944


\bibitem{Cardy}
J. L. Cardy,
Conformal invariance and statistical mechanics,
Lectures given at Les Houches (1988), in: Fields, strings and
critical phenomena, eds. E. Br\'ezin, J. Zinn-Justin
(North-Holland, Amsterdam, 1989)

\bibitem{Ginsparg}
P. Ginsparg,
Applied conformal field theory,
Lectures given at Les Houches (1988), in: Fields, strings and
critical phenomena, eds. E. Br\'ezin, J. Zinn-Justin
(North-Holland, Amsterdam, 1989)


\bibitem{QFTbI}
K. Symanzik,
Nucl. Phys. B190 [FS3] (1981) 1

\bibitem{QFTbII}
M. L\"uscher,
Nucl. Phys. B254 (1985) 52


\bibitem{SommerScaleA}
R. Sommer,
Nucl. Phys. B411 (1994) 839



\bibitem{Shin}
D.-S. Shin,
Nucl. Phys. B525 (1998) 457


\bibitem{MichaelPerantonis}
S. Perantonis, C. Michael,
Nucl. Phys. B347 (1990) 854

\bibitem{KutiEtAlI}
K. J. Juge, J. Kuti, C. J. Morningstar,
Nucl. Phys. B (Proc.Suppl.) 63 (1998) 326;
{\it ibid.}\/ 73 (1999) 590;
{\it ibid.}\/ 83 (2000) 503;
{\it ibid.}\/ 106 (2002) 691

\bibitem{KutiEtAlII}
K. J. Juge, J. Kuti, C. J. Morningstar,
From surface roughening to QCD string theory,
in: Non-perturbative QFT methods and their applications
(Budapest 2000),
eds. Z. Horv\'ath, E. Palla
(World Scientific, Singapore, 2001)


\bibitem{Michael}
C. Michael,
Glueballs, hybrid and exotic mesons, and string breaking,
in: Quark confinement and the hadron spectrum IV (Vienna 2000),
eds. W. Lucha, K. M. Maung (World Scientific, Singapore, 2002)


\bibitem{Teper}
M. Teper,
Phys. Rev. D59 (1999) 014512


\bibitem{Fischler}
W. Fischler, 
Nucl. Phys. B129 (1977) 157

\bibitem{Billoire}
A. Billoire,
Phys. Lett. B92 (1980) 343

\bibitem{Peter}
M. Peter, 
Nucl. Phys. B501 (1997) 471

\bibitem{Schroder}
Y. Schr\"oder, 
Phys. Lett. B447 (1999) 321;
DESY-THESIS-1999-021

\bibitem{Melles}
M. Melles,
Phys. Rev. D62 (2000) 074019


\bibitem{AlphaI}
M. L\"uscher, R. Sommer, P. Weisz, U. Wolff,
Nucl. Phys. B413 (1994) 481

\bibitem{AlphaII}
S. Capitani, M. L\"uscher, R. Sommer, H. Wittig,
Nucl. Phys. B544 (1999) 669


\bibitem{SilviaRainerII}
S. Necco, R. Sommer,
Phys. Lett. B523 (2001) 135


\bibitem{Symanzik}
K. Symanzik,
Nucl. Phys. B226 (1983) 187, 205


\bibitem{Polyakov}
A. M. Polyakov,
Int. J. Mod. Phys. A14 (1999) 645

\endbibliography

\bye